\documentclass[12pt]{article}
\usepackage[left=1in,top=1in,right=1in,bottom=1in,head=.1in,nofoot]{geometry}
\usepackage{setspace,url,bm,amsmath} % For double-spacing, URL font, math symbols

\usepackage{titlesec} % Section header formatting
\titlelabel{\thetitle.\quad} % Section header formatting
\titleformat*{\section}{\bf\large\center\uppercase} % Section header formatting

\usepackage{amssymb}
\usepackage{amsmath}
\usepackage{MnSymbol}
\usepackage{graphicx}
\usepackage{booktabs}
\usepackage[margin=20pt]{subcaption}
\usepackage{setspace}
\usepackage{blindtext}
\usepackage[utf8]{inputenc}
\usepackage{tabularx, ragged2e}
\usepackage{bbm}
\usepackage{latexsym}
\usepackage{caption}
\usepackage{hyperref}
\usepackage{float}
\usepackage{comment}
\usepackage{dcolumn}
\usepackage{adjustbox}

\newcolumntype{Y}{>{\centering\arraybackslash}X}

\begin{document}

\centerline{ \large \textbf{Improving Covariate Balance in $2^K$ Factorial Designs via Rerandomization}}
\vspace{2pc}
\centerline{Zach Branson, Tirthankar Dasgupta, and Donald B. Rubin}
\centerline{Harvard University, Cambridge, USA}

\begin{center}
\textbf{Abstract}
\justify
Factorial designs are widely used in agriculture, engineering, and the social sciences to study the causal effects of several factors simultaneously on a response. The objective of such a design is to estimate all factorial effects of interest, which typically include main effects and interactions among factors. To estimate factorial effects with high precision when a large number of pre-treatment covariates are present, balance among covariates across treatment groups should be ensured. We propose utilizing rerandomization to ensure covariate balance in factorial designs. Although both factorial designs and rerandomization have been discussed before, the combination has not. Here, theoretical properties of rerandomization for factorial designs are established, and empirical results are explored using an application from the New York Department of Education.
\end{center}

% \setcounter{tocdepth}{4}
% 1 Section
% 2 Subsection
% 3 Subsubsection
% 4 Paragraph
% 5 Subparagraph

\newpage

\section{Introduction}

$2^K$ factorial designs involve $K$ factors each with two levels, often denoted as the ``high level'' and ``low level'' of the factor (Yates 1937, Fisher 1942). With $K$ factors, there are $2^K$ unique treatment combinations to which units can be assigned, and often the same number of units are assigned to each combination. Factorial designs are often discussed in an industrial setting, where units are essentially identical and the assignment of units to treatments is arbitrary. However, in recent years factorial designs have become more prevelant in clinical trials and the social sciences, where pre-treatment covariates are available and reveal that units differ. For example, the New York Department of Education (NYDE) had five ``incentive programs'' to introduce to high schools, and it wanted to estimate the effect of these programs and their combinations on schools' performance. Given 50 pre-treatment covariates for each of the 1,376 schools, how should the department allocate the schools to the 32 different treatment combinations of the design such that the effects of the incentive programs and their combinations are well-estimated?

An initial idea for this example is to randomize the schools to the 32 treatment combinations. Randomized experiments are considered the ``gold standard" because randomization balances all potential confounders \textit{on average} (Krause and Howard 2003, Morgan and Rubin 2012). However, many have noted that randomized experiments can yield ``bad allocations,'' where some covariates are not well-balanced across treatment groups (Seidenfeld 1981, Lindley 1982, Papineau 1994, and Rosenberger and Sverdlov 2008). Covariate imbalance among different treatment groups complicates the interpretation of estimated treatment effects.

If ``bad allocations" are a concern for treatment-versus-control experiments, they are even more of a concern for factorial designs, because any randomization may create covariate imbalance across some of the $2^K$ treatment combinations. This point has been given little attention in the literature. Classic experimental design textbooks like Box, Hunter, and Hunter (2005) and Wu and Hamada (2009) suggest using blocking to balance important covariates; however, the use of blocking is not obvious with many covariates, some with many levels. Additionally, Wu and Hamada (2009) note that blocking can increase precision for some factorial effect estimators and ``sacrifice" the precision of other factorial effect estimators. To address this issue, we propose a rerandomization algorithm for balanced $2^K$ factorial designs based on Morgan and Rubin (2012), which developed a framework for rerandomization in the treatment-versus-control case. Here we establish several theoretical properties of rerandomization in balanced $2^K$ factorial designs that increase the precision of factorial effect estimators.

Both rerandomization and factorial designs have been explored since Fisher in the 1920s. To our knowledge, however, no one has laid out the framework for implementing rerandomization for factorial designs. Rubin (2008) noted that many did not implement rerandomization because it was computationally intensive; however, with recent improvements in computational power, some have revisited rerandomization. For example, Cox (2009), Bruhn and McKenzie (2009), and Worrall (2010) all discuss and recommend rerandomization, and Morgan and Rubin (2012) formalized these recommendations in treatment-versus-control settings.

Also, often there are few pre-experiment covariates to consider in a factorial design, or they are categorical - such as the ``batch" of units produced, as described in Box, Hunter, and Hunter (2005) - and thus simple blocking is an appropriate strategy. In contrast, Wald, et al. (1991), Apfel, et al. (2002), and Bays et, al. (2004) all describe clinical trials that utilized randomized factorial designs with non-categorical covariates, which could have benefited from a design that ensured covariate balance. To illustrate how rerandomization can be utilized in such situations, we use an education example discussed in Dasgupta, et al. (2015).

Our proposed rerandomization algorithm is not the first procedure that attempts to balance non-categorical covariates for experiments with multiple treatments. The Finite Selection Model (FSM) developed by Morris (1979) assigns units to multiple treatment groups such that covariates are relatively balanced among the groups. Morgan and Rubin (2012) noted that rerandomization and the FSM both attempt to ensure covariate balance, but the FSM does not maintain the correlation structure among covariates, whereas rerandomization can.

Xu and Kalbfleisch (2013) proposed the ``balance match weighted design'' for multiple treatment groups, which performs many randomizations and then selects the randomization that yields the best covariate balance. This is similar to rerandomization, but rerandomization's theoretical guarantees, such as balancing on unobserved covariates on average in addition to improving balance for observed covariates, is appealing. Our rerandomization algorithm can also incorporate various desiderata, such as factorial effects and covariates that vary in importance, which makes the procedure particularly flexible.

In Section \ref{s:review} we review rerandomization for the treatment-versus-control case, and in Section \ref{s:2K} we establish notation for $2^K$ factorial designs using the potential outcomes framework. In Section \ref{ss:obs} we outline the proposed rerandomization procedure, and in Section \ref{s:properties} we establish theoretical properties that formalize the ways rerandomization is preferable to standard randomization. In Section \ref{s:schools} we use our rerandomization procedure on data from the NYDE.

\section{Review of Rerandomization} \label{s:review}

Rubin (2008) recalled a conversation with Bill Cochran, who in turn recalled a conversation with R.A. Fisher, who asserted that a way to protect ourselves against particularly bad randomizations is to rerandomize until a randomization is ``acceptable.'' Morgan and Rubin (2012) suggested implementing rerandomization for a treatment-versus-control experiment as follows:

\begin{enumerate}
\item Collect covariate data.
\item Specify a balance criterion determining when a randomization is acceptable.
\item Randomize units to treatment and control groups.
\item Check the balance criterion. If the criterion is met, go to Step 5. Otherwise, return to Step 3.
\item Conduct the experiment using the final randomization obtained in Step 4.
\item Analyze the results using a randomization test, keeping only simulated randomizations that satisfy the balance criteria specified in Step 2.
\end{enumerate}

Morgan and Rubin (2012) used the squared Mahalanobis distance (Mahalanobis 1936) as a measure for covariate balance. With $n$ units, half assigned to treatment and half assigned to control, and $p$ observed covariates for each unit, the squared Mahalanobis distance for the treatment-versus-control situation is defined as:
\begin{align*}
M \equiv (\bar{{\bm{x}}}_T - \bar{{\bm{x}}}_C)^T \text{cov}[(\bar{{\bm{x}}}_T - \bar{{\bm{x}}}_C)]^{-1}(\bar{{\bm{x}}}_T - \bar{{\bm{x}}}_C),
\end{align*}
where $\bar{{\bm{x}}}_T$ is the $p$-component column vector of covariate means for units assigned to the treatment and $\bar{{\bm{x}}}_C$ is analogously defined for the control. A randomization is declared acceptable if $M \leq a$ for some threshold $a$. The Mahalanobis distance is well-known within the matching and observational study literature, where it is used to find subsets of the treatment and control that are similar (Rubin 1976, Rosenbaum and Rubin 1985, Gu and Rosenbaum 1993, Rubin and Thomas 2000). Constraining $M \leq a$ can be viewed as finding allocations where the treatment and control covariate means are ``similar enough,'' where the ``enough'' is determined by the threshold $a$. Morgan and Rubin (2012) note that - similar to Rosenbaum and Rubin's (1985) argument that matching using the Mahalanobis distance reduces bias due to imbalances in covariates from observational studies - rerandomization using $M$ reduces the sampling \textit{variance} of the standard treatment effect estimator when outcome variables are correlated with covariates.

Morgan and Rubin (2012) showed that $M$ closely follows a chi-squared distribution with degrees of freedom equal to $p$. Thus, $a$ can be selected by first deciding the percentage, $p_a$, of randomizations that will be ``acceptably well-balanced,'' and then setting $a$ to the $p_a$th percentile of the $\chi^2_p$ distribution. For example, if there are five covariates, and we want to select from the 1\% ``most balanced'' randomizations, then $a$ is set equal to the first percentile of the $\chi^2_5$ distribution.

Morgan and Rubin (2012) mention two options for balancing covariates among multiple treatment groups:
\begin{enumerate}
\item Create a criterion for each pairwise comparison among the treatment groups, and then rerandomize if any group does not satisfy the criterion.
\item Use a statistic that measures multivariate balance, such as those used in standard MANOVA analyses.
\end{enumerate}
To implement (1), a criterion for  each ${2^K \choose 2} = 2^{K-1}(2^K - 1)$ pairwise comparison must be chosen, which may be computationally burdensome. To implement (2), there must be a notion of ``within-group" variance, which is not immediate for unreplicated $2^K$ factorial designs where only one unit is assigned to each treatment combination. Furthermore, we may not want to estimate all factorial effects with the same level of precision; for instance, typically we want to estimate main effects more precisely than high-order interactions, and it is not clear how to incorporate this desideratum into (1) or (2). We propose an intuitive adjustment to (1) for balanced $2^K$ factorial designs, which is equivalent to (2) for replicated factorial designs. The proposed adjustment also allows hierarchies of effect importance.

\section{Notation for $2^K$ Designs under the Potential Outcomes Framework} \label{s:2K}

Consider a balanced $2^K$ factorial design with $n = r2^K$ units and $r$ replicates assigned to each of the $2^K$ treatment combinations. In a $2^K$ factorial design there are $2^K -1$ factorial effects: $K$ main effects, $\binom{K}{2}$ two-way interactions, $\binom{K}{3}$ three-way interactions, and so on.

The $2^K$ treatment combinations of a $2^K$ factorial design are often arranged in a specific order and represented as a $2^K \times K$ matrix whose elements are either -1 (representing the ``low level'' of a factor) or +1 (representing the ``high level'' of a factor), and thus each row indicates a unique combination of factor assignments. This matrix is often referred to as the \textit{design matrix} (Wu and Hamada 2009). One such order for the -1s and +1s is the lexicographic order (Espinosa, et al. 2015) in which each column starts with $-1$, making the first row of the matrix a $K$-component vector of -1s. In the first column, the sign is switched to $+1$ for the second half (i.e., $2^{K-1}$) of the components. In the second column, the sign is switched after every one-fourth (i.e., $2^{K-2}$) of the components. Proceeding this way, the last column consists of alternating -1s and +1s. We denote the design matrix by $\mathbf{G}$; see Table \ref{tab:designMatrix} for a $2^3$ factorial design using the lexicographic order. Another well-known order is Yates' order, in which the columns appear in exactly the reverse order of the lexicographic order.

\begin{table}
\centering
\caption{The Design Matrix, $\mathbf{G}$, for a $2^3$ design} \label{tab:designMatrix}
\begin{tabularx}{0.25\textwidth}{|Y|Y|Y|}
\hline
 $A$ & $B$ & $C$ \\ \hline
  -1 & -1 & -1   \\
  -1 & -1 & +1   \\
  -1 & +1 & -1   \\
  -1 & +1 & +1   \\
  +1 & -1 & -1   \\
  +1 & -1 & +1   \\
  +1 & +1 & -1  \\
  +1 & +1 & +1   \\ \hline
\end{tabularx}
\end{table}

\begin{table}
\centering
\caption{$\widetilde{\mathbf{G}}$ for a $2^3$ design (columns 2-4 represent the design matrix $\mathbf{G}$)} \label{tab:matrix}
\begin{tabularx}{0.75\textwidth}{|c||*3{Y}||*4{Y}|}
\hline
Mean Column & \multicolumn{3}{c||}{Main effect columns} & \multicolumn{4}{c|}{Interaction columns} \\
&  $A$ & $B$ & $C$ & $AB$ & $AC$ & $BC$ & $ABC$ \\ \hline
+1 & -1 & -1 & -1 & +1 & +1 & +1 & -1 \\
+1 & -1 & -1 & +1 & +1 & -1 & -1 & +1 \\
+1 & -1 & +1 & -1 & -1  & +1 & -1 & +1 \\
+1 & -1 & +1 & +1 & -1  & -1 & +1 & -1 \\
+1 & +1 & -1 & -1 &  -1 & -1 & +1 & +1 \\
+1 & +1 & -1 & +1 &  -1 & +1 & -1 & -1 \\
+1 & +1 & +1 & -1 &  +1 & -1 & -1 & -1 \\
+1 & +1 & +1 & +1 &  +1 & +1 & +1 & +1 \\ \hline
\end{tabularx}
\end{table}

To define the interaction effects, we expand $\mathbf{G}$ (the columns labeled ``main effect columns'' in Table \ref{tab:matrix}) by augmenting its columns. The column for a specific interaction is created using component-wise multiplication of the corresponding main-effects columns. For example, the last column in Table \ref{tab:matrix} represents the three-way interaction among factors $A$, $B$ and $C$, and is obtained by multiplying the components in the three columns of $\mathbf{G}$. Having appended the interaction columns to the right of $\mathbf{G}$ (columns 5-8 of Table \ref{tab:matrix}) to define the interaction effects, a column of +1s is appended to the left of $\mathbf{G}$ (first column of Table \ref{tab:matrix}) which defines the mean effect. The result is a $2^K \times 2^K$ matrix, denoted by $\widetilde{\mathbf{G}}$. The rows of $\widetilde{\mathbf{G}}$ are indexed by $j = 1, \dots, 2^K$, one row for each treatment combination, as indicated by $\mathbf{G}$, and the columns are indexed by $f = 0, 1, \dots, 2^K-1$; ``$f$'' for factorial effects. Let $\widetilde{\mathbf{G}}_j.$ and $\widetilde{\mathbf{G}}_{.f}$ denote the $j$th row and $f$th column of $\widetilde{\mathbf{G}}$, respectively.

Let $Y_i(j)$, $i=1, \ldots, n$, $j = 1, \ldots, 2^K$ denote the potential outcome for the $i$th unit when exposed to the $j$th treatment combination, and let ${\bm Y}_i = \left(Y_i(1), \ldots, Y_i(2^K) \right)$ denote the row vector of the $2^K$ potential outcomes for unit $i$. The $i$th row of the left part of Table \ref{tab:fac_effects} shows ${\bm Y}_i$ for a $2^3$ design.

\begin{table}
\centering
\caption{Unit-level and Population-Level Factorial Effects for a $2^3$ Design} \label{tab:fac_effects}
\begin{adjustbox}{max width=\textwidth}
\begin{tabular}{ccccc}
\hline
Unit ($i$) & Potential outcomes (${\bm Y}_i)$ & Mean of unit $i$ ($\theta_{i0}$) & Factorial effect $\theta_{i,f}$ \\ \hline
1 & ${\bm Y}_1 = \left( Y_1(1), \cdots, Y_1(8) \right)$  & $ 8^{-1} {\bm Y}_1 \widetilde{\mathbf{G}}_{.0} $  &  $ 4^{-1} {\bm Y}_1 \widetilde{\mathbf{G}}_{.f}  $ \\
2 & ${\bm Y}_2 = \left( Y_2(1), \cdots, Y_2(8) \right)$  & $ 8^{-1} {\bm Y}_2 \widetilde{\mathbf{G}}_{.0} $  &  $ 4^{-1} {\bm Y}_2 \widetilde{\mathbf{G}}_{.f}$ \\
$\vdots$ & $\vdots$ & $\vdots$ & $\vdots$ \\
$n$ & ${\bm Y}_n = \left( Y_n(1), \cdots, Y_n(8) \right)$    &  $ 8^{-1} {\bm Y}_n \widetilde{\mathbf{G}}_{.0} $  & $ 4^{-1} {\bm Y}_n \widetilde{\mathbf{G}}_{.f}  $ \\ \hline
Average & $\bar{\bm Y} = n^{-1} \left( \sum_i Y_i(1), \cdots, \sum_i Y_i(8) \right)$ & $ 8^{-1} \bar{\bm Y} \widetilde{\mathbf{G}}_{.0} $ & $\bar{\theta}_f =  4^{-1} \bar{\bm Y} \widetilde{\mathbf{G}}_{.f}$ \\ \hline
\end{tabular}
\end{adjustbox}
\end{table}

Following Dasgupta, et al. (2015), the $f$th linear factorial effect for unit $i$ is:
\begin{align*}
\theta_{if} = \frac{1}{2^{K-1}} {\bm Y}_i \widetilde{\mathbf{G}}_{.f}, \;\ i=1, \ldots, n, \;\ f=1, \ldots, 2^K-1
\end{align*}
and the population-level $f$th factorial-effect is defined as:
\begin{equation}
\bar{\theta}_f = \frac{1}{n} \sum_{i=1}^n \theta_{if}. \label{eq:pop_fac}
\end{equation}
The $f$th factorial effect at the unit level and the population level, represented as functions of the potential outcomes, are shown in the last column of Table \ref{tab:fac_effects}. The second-to-last column of Table \ref{tab:fac_effects} shows the unit-level mean of the potential outcomes
\begin{align*}
	\theta_{i0} = \frac{1}{2^K} {\bm Y}_i \widetilde{\mathbf{G}}_{.0}
\end{align*}
and their grand mean $\bar{\theta}_0$. The population-level grand mean $\bar{\theta}_0$ and the linear factorial effects $\bar{\theta}_1, \ldots, \bar{\theta}_{2^K-1}$ are the estimands (objects of interest) in the standard linear finite-population framework described here. They need to be estimated because only one element of ${\bm Y}_i$ can be observed for each $i$. We discuss unbiased estimators of these estimands in Section \ref{ss:obs}.

The vector $(\theta_{i0}, \ldots, \theta_{i(2^K-1)})$ of estimands for unit $i$ is a linear transformation of the vector ${\bm Y}_i$ of potential outcomes. Letting the factorial effects vector for unit $i$ be
\begin{equation}
{\bm \theta}_i =  \left( \theta_{i0}, \frac{ \theta_{i1}}{2}, \ldots, \frac{\theta_{i(2^K-1)}}{2} \right), \hspace{0.1 in} i = 1, \dots, n, \label{eq:thetaVector}
\end{equation}
straightforward algebra shows that the potential outcomes for unit $i$ can be written as
\begin{align*}
{\bm Y}_i = {\bm \theta}_{i.} \widetilde{\mathbf{G}}^T
\end{align*}
so the $j$th component of ${\bm Y}_i$ is
\begin{equation}
Y_i(j) = {\bm \theta}_i \widetilde{\mathbf{G}}_{j.}^T , \label{eq:model1a}
\end{equation}

\section{The assignment mechanism, unbiased estimation of factorial effects, and the rerandomization algorithm} \label{ss:obs}

Randomized balanced $2^K$ factorial designs assign $n = r2^K$ units to one of $2^K$ treatment combinations with equal probability such that $r$ units are assigned to each combination. Each combination corresponds to a row of the design matrix $\mathbf{G}$. Let ${\bm{W}}$ be a $n \times K$ random matrix where the $i$th row of ${\bm{W}}$, ${\bm{W}}_{i.}$, indicates the treatment assignment for unit $i$, and has probability $\frac{1}{2^K}$ of being the $j$th row of $\mathbf{G}$: $P(\bm{W}_{i.} = \mathbf{G}_{j.}) = \frac{1}{2^K}$ for $i = 1, \dots, n$, $j = 1, \dots, 2^K$. For notational convenience, we expand ${\bm{W}}$ to $\widetilde{\bm{W}}$ such that $P(\widetilde{\bm{W}}_{i.} = \widetilde{\mathbf{G}}_{j.}) = \frac{1}{2^K}$, where the first column of $\widetilde{\bm{W}}$, $\widetilde{\bm{W}}_{.0}$, is not stochastic and is +1s, as in $\widetilde{\mathbf{G}}$; every other element of $\widetilde{\bm{W}}$ for $i = 1, \dots, n$ and $f \in F \equiv \{1, \dots, 2^K-1 \}$ is defined as
\begin{align}
\widetilde{W}_{i f} = \begin{cases} +1 &\mbox{if the $i$th unit is assigned to high level of $f$} \\
-1 & \mbox{if the $i$th unit is assigned to low level of $f$} \end{cases} \label{eq:Wf}
\end{align}
Let $\widetilde{\bm{W}}_{.f}$ be the $n$ x $1$ column vector denoting the assigned level of some $f \in F$ for all units. A particular random allocation of units in a $2^K$ design corresponds to one realization of $\widetilde{\bm{W}}$, the observed one, $\widetilde{\bm{W}}^{\text{obs}}$. The observed outcome for the $i$th unit will be the potential outcome $Y_i(j)$ when $\widetilde{\bm{W}}^{\text{obs}}_{i.} = \widetilde{\bm{G}}_{j.}$. Let ${\bm y}_{\text{obs}}$ be the $n$-component column vector of observed outcomes for the $n$ units. The standard estimator of the factorial effect $\bar{\theta}_f$ defined in (\ref{eq:pop_fac}) can be written in terms of the observed outcomes and $\widetilde{\bm{W}}$:
\begin{align}
	\hat{\theta}_f = \bar{y}_{f^+} - \bar{y}_{f^-} = \frac{{\bm{y}}_{\text{obs}}^T \widetilde{\bm{W}}_{.f}}{n/2} \label{eq:factorialEffectEstimator}
\end{align}
where $\bar{y}_{f^+}$ is the mean outcome for units assigned to the high level of some $f \in F$, and $\bar{y}_{f^-}$ is analogously defined for the low level of $f$.

Rerandomization involves randomizing until an allocation is declared ``acceptable,'' using an acceptance criterion $\phi({\mathbf{X},\widetilde{\bm{W}}})$, where $\mathbf{X}$ is the $n \times p$ covariate matrix, and $\phi$ equals one if an allocation is ``acceptable'' and zero otherwise.

Consider an acceptance criterion that is symmetric in $\widetilde{{\bm{W}}}$, i.e., a $\phi$ such that $\phi({\mathbf{X},\widetilde{\bm{W}}}) = \phi({\mathbf{X}, -\widetilde{\bm{W}}})$. Theorem 1 below establishes that the standard factorial effect estimators are unbiased under any rerandomization scheme that uses a symmetric acceptance criterion.

\noindent
\textbf{Theorem 1}: Suppose a completely randomized balanced $2^K$ factorial design is rerandomized when $\phi({\mathbf{X},\widetilde{\bm{W}}}) = 0$ for some symmetric acceptance criterion. Then, for all $f \in F$,
\begin{align*}
	\mathbb{E}[\hat{\theta}_f | \phi({\mathbf{X},\widetilde{\bm{W}}}) = 1] = \bar{\theta}_f,
\end{align*}
where $\hat{\theta}_f$ is the estimator defined in (\ref{eq:factorialEffectEstimator}) and $\bar{\theta}_f$ is the population-level estimand defined in (\ref{eq:pop_fac}). Because $\phi$ is symmetric in $\widetilde{\bm W}$, the proof of the unbiasedness of $\hat{\theta}_f$ under rerandomization is analogous to that in Morgan and Rubin (2012) for the treatment-versus-control situation.

If the potential outcomes are correlated with pre-experiment covariates, then so will be the observed outcomes and the estimator $\hat{\theta}_f$ for any $f \in F$. Intuitively, we can increase the precision of $\hat{\theta}_f$ by ensuring covariates are ``well-balanced'' over the two groups used to calculate $\hat{\theta}_f$: the ``treatment'' (units assigned to the high level of $f$) and the ``control'' (units assigned to the low level), which suggests a balance function that measures the covariate balance between all pairs of these groups.

One such balance function is the squared Mahalanobis distance proposed by Morgan and Rubin (2012). A way to measure the covariate balance between the ``treatment'' and ``control'' for a particular $f$ is to define
\begin{align}
M_f &\equiv (\bar{{\bm{x}}}_{f^+} - \bar{{\bm{x}}}_{f^-})^T \text{cov}[(\bar{{\bm{x}}}_{f^+} - \bar{{\bm{x}}}_{f^-})]^{-1} (\bar{{\bm{x}}}_{f^+} - \bar{{\bm{x}}}_{f^-}) \notag \\
&= \frac{n}{4} (\bar{{\bm{x}}}_{f^+} -\bar{{\bm{x}}}_{f^-})^T \text{cov}[\mathbf{X}]^{-1}(\bar{{\bm{x}}}_{f^+} - \bar{{\bm{x}}}_{f^-}) \label{eq:mahalanobisDistance}
\end{align}
where $\bar{{\bm{x}}}_{f^+}$ is the $p$-component vector of covariate means for units assigned to the high level of $f$ and $\bar{{\bm{x}}}_{f^-}$ is analogously defined. Note that, analogous to (\ref{eq:factorialEffectEstimator}), $\bar{{\bm{x}}}_{f^+} - \bar{{\bm{x}}}_{f^-} = \frac{\mathbf{X}^T \widetilde{\bm W}_{.f}}{n/2}$.

The covariate balance between the ``treatment'' and the ``control'' for a particular $f$ is declared ``acceptable'' by the acceptance criterion
\begin{align}
	\phi_f({\mathbf{X}, \widetilde{\bm{W}}}) = \begin{cases}
	1 &\mbox{if } M_f \leq a \\
	0 &\mbox{if } M_f > a
	\end{cases} \label{eq:phif}
\end{align}
for a predetermined threshold $a$. An intuitive procedure that parallels Morgan and Rubin (2012) is to randomize until $\phi_f({\mathbf{X}, \widetilde{\bm{W}}}) = 1$ in order to increase the covariate balance between the ``treatment'' and ``control'' for a particular $f$. We can do this for every $f \in F$, and thereby define the overall acceptance criterion as
\begin{align}
	\phi({\mathbf{X},\widetilde{\bm{W}}}) = \prod_{f \in F} \phi_f({\mathbf{X}, \widetilde{\bm{W}}}) = \begin{cases}
	1 &\mbox{if } \max_{f \in F} M_f \leq a \\
	0 &\mbox{if } \max_{f \in F} M_f > a \label{eq:phi}
	\end{cases}
\end{align}
We thus propose the following rerandomization procedure for balanced $2^K$ factorial designs:
\begin{enumerate}
	\item Create a squared Mahalanobis distance criterion $M_f$ for each $f \in F$.
	\item Choose a threshold criterion $a$ as in Morgan and Rubin (2012).
	\item Randomize until $\phi({\mathbf{X},\widetilde{\bm{W}}}) = 1$, where $\phi$ is defined as in (\ref{eq:phi}).
\end{enumerate}

\noindent
We have the following corollary:

\noindent
\textbf{Corollary 1}: Theorem 1 holds if $\phi({\mathbf{X},\widetilde{\bm{W}}})$ is defined as in (\ref{eq:phi}).

Section \ref{s:properties} establishes that the above rerandomization algorithm increases the precision of all factorial effect estimators compared to pure randomization.

\section{Precision Properties of Rerandomization} \label{s:properties}

The proposed rerandomization algorithm checks $M_f$ for \textit{all} $f \in F$, i.e., $\phi({\mathbf{X},\widetilde{\bm{W}}}) = 1$ iff $\phi_f({\mathbf{X}, \widetilde{\bm{W}}}) = 1$ for all $f \in F$. Thus, both the marginal and joint distributions of $\{\bar{\mathbf{x}}_{f^+} - \bar{\mathbf{x}}_{f^-} : f \in F\}$ and $\{\hat{\theta}_f : f \in F\}$ need to be examined.

\noindent
\textbf{Theorem 2}: Assume a completely randomized balanced $2^K$ factorial design is rerandomized using the algorithm proposed at the end of Section \ref{ss:obs}. Then,
\begin{align*}
\mathbb{E}[\overline{{\bm{x}}}_{f^+} - \overline{{\bm{x}}}_{f^-} | \phi({\mathbf{X},\widetilde{\bm{W}}}) = 1] = 0
\end{align*}
The proof of Theorem 2 follows immediately by symmetry of the acceptance criterion, as in Morgan and Rubin (2012).

\noindent
\textbf{Lemma 1}: Assume a completely randomized balanced $2^K$ factorial design is rerandomized using the algorithm proposed at the end of Section \ref{ss:obs}, and the covariate means are multivariate normal. Then, the elements of $\{\phi_f({\mathbf{X}, \widetilde{\bm{W}}}): f \in F\}$ defined in (\ref{eq:phif}) are mutually independent.

The proof of Lemma 1 is in the Appendix.

\noindent
\textbf{Theorem 3}: Assume a completely randomized balanced $2^K$ factorial design is rerandomized using the algorithm proposed at the end of Section \ref{ss:obs}, and the covariate means are multivariate normal. Then:

\noindent
First, for all $f \in F$,
\begin{align*}
\text{cov}[\overline{{\bm{x}}}_{f^+} - \overline{{\bm{x}}}_{f^-} | \phi({\mathbf{X},\widetilde{\bm{W}}}) = 1]  = v_a \text{cov}[\overline{{\bm{x}}}_{f^+} - \overline{{\bm{x}}}_{f^-}]
\end{align*}
where
\begin{align}
v_a = \frac{2}{p} \frac{\gamma ( \frac{p}{2} + 1, \frac{a}{2})}{\gamma(\frac{p}{2}, \frac{a}{2})}, \label{eq:vaDef}
\end{align}
and $\gamma$ is the incomplete gamma function $\gamma(b, c) \equiv \int_0^c y^{b-1} e^{-y} dy$.

\noindent
And second, for $f_1, f_2 \in F, f_1 \neq f_2$,
\begin{align*}
\text{cov}[\overline{{\bm{x}}}_{f_1^+} - \overline{{\bm{x}}}_{f_1^-}, \overline{{\bm{x}}}_{f_2^+} - \overline{{\bm{x}}}_{f_2^-} | \phi({\mathbf{X},\widetilde{\bm{W}}}) = 1] = \mathbf{0}
\end{align*}
Theorem 3 is proved in the Appendix.

Theorems 2 and 3 establish that rerandomization leads to unbiased estimators and reduces the variance of ($\bar{{\bm x}}_{f^+} - \bar{{\bm x}}_{f^-})$, and that this reduction is the same for all covariates. We define the \textit{percent reduction in variance} for any covariate $p$ and $f \in F$ as:
\begin{align}
100 \left( \frac{\text{var} [\bar{x}_{p, f^+} - \bar{x}_{p, f^-} ] - \text{var}[\overline{x}_{p, f^+} - \bar{x}_{p, f^-} | \phi({\mathbf{X},\widetilde{\bm{W}}}) = 1]}{\text{var}[\bar{x}_{p,f^+} - \bar{x}_{p,f^-} ]}\right) = 100(1 - v_a) \label{eq:va}
\end{align}
Therefore, for any covariate $p$ and $f \in F$, the rerandomization algorithm will reduce the variance of $(\bar{x}_{p, f^+} - \bar{x}_{p, f^-})$ in expectation by $100(1-v_a) \%$, compared to pure randomization.

To state properties of the marginal and joint distributions of the factorial effect estimators $\{\hat{\theta}_f: f \in F\}$, assumptions must be made about the relationship between the potential outcomes and the factorial effects and covariates. Suppose the factorial effects ${\bm \theta}_i$ defined in (\ref{eq:thetaVector}) are constant across units and there is no interaction between factorial effects and covariate effects. Then, the potential outcomes can be written using the following linear model:
\begin{equation}
Y_i(j) = {\bm \theta}_i \widetilde{\mathbf{G}}_{j.}^T + {\bm{x}}_i {\bm \beta} + \epsilon_i, \hspace{0.1 in} i = 1, \dots, n, j = 1, \dots, 2^K \label{eq:model1b}
\end{equation}
where $\widetilde{\mathbf{G}}_{j.}$ is the $j$th row of $\widetilde{\mathbf{G}}$ defined in Section \ref{s:2K}, ${\bm \beta}$ is the $p$-component column vector of fixed covariate coefficients, and $\epsilon_i$ indicates any deviations from the linear model. Then, the standard unbiased estimator (\ref{eq:factorialEffectEstimator}) can be written as:
\begin{align}
 	\hat{\theta}_f = \bar{\theta}_f + \boldsymbol{\beta}^T (\bar{{\bm{x}}}_{f^+} - \bar{{\bm{x}}}_{f^-}) + (\bar{{\bm{\epsilon}}}_{f^+} - \bar{{\bm{\epsilon}}}_{f^-}) \label{eq:linearModel}
 \end{align}
and the theorem below follows.

\noindent
\textbf{Theorem 4}: Assume (a) a completely randomized balanced $2^K$ factorial design is rerandomized using the algorithm proposed at the end of Section \ref{ss:obs}, (b) the covariate means are multivariate normal, (c) factorial effects are constant across units, and (d) there is no interaction between factorial effects and covariate effects. Then, for all $f \in F$,
\begin{align}
\text{var}(\hat{\theta}_f | \phi({\mathbf{X},\widetilde{\bm{W}}}) = 1) = \left(1 - (1 - v_a)R^2) \text{var}(\hat{\theta}_f \right) \label{eq:rerandomizationBenefit}
\end{align}
and for $f_1, f_2 \in F$, such that $f_1 \neq f_2$,
\begin{align*}
	\text{cov}(\hat{\theta}_{f_1}, \hat{\theta}_{f_2} | \phi({\mathbf{X},\widetilde{\bm{W}}}) = 1) = 0
\end{align*}
where $R^2$ is the squared multiple correlation coefficient between $\mathbf{y_{\text{obs}}}$ and $\mathbf{X}$, and $v_a$ is defined in (\ref{eq:vaDef}). The proof of Theorem 4 is in the Appendix.

Theorem 4 has several implications. First, randomizing until $M_f \leq a$ for all $f \in F$ on average increases the precision of all factorial effect estimators equally by $100(1-v_a)R^2$ percent. Likewise, for some subset $F^* \subset F$, randomizing until $M_f \leq a$ equally increases the precision of $\hat{\theta}_f$ for all $f \in F^*$, again by $100(1-v_a)R^2$ percent, but this does not affect the precision of $\hat{\theta}_f$ for any $f \notin F^*$, by the uncorrelated result of Theorem 4. Furthermore, different thresholds can be chosen for each squared Mahalanobis distance criterion. For example, one can randomize until $M_f \leq a_f$, where $a_f$ differs across $f$. Choosing a smaller $a_f$ for each $f \in F^*$ ensures a higher increase in precision for the corresponding factorial effect estimator $\hat{\theta}_f$.

Thus, we can adapt our rerandomization procedure according to tiers of importance for factorial effects. Furthermore, we can do the same for covariates, analogous to Morgan and Rubin (2015), which shows how to adapt rerandomization according to tiers of importance for covariates in the treatment-versus-control case. 

To conduct inference using rerandomization, the significance levels of hypotheses should be calculated using a permutation test (Fisher 1942). However, during the permutation test, the distribution of the test statistic under Fisher's sharp null must be created using randomizations that would be accepted under rerandomization (Morgan and Rubin 2012). Corrections for multiple testing and selection of active versus inactive effects (as in Espinosa, et al. 2015) are topics for future work. 

Theorems 3 and 4 require $n$ to be sufficiently large such that the covariate means are multivariate normal. If the covariate means are multivariate normal, then the Mahalanobis distance is $\chi^2_p$ (Mardia et al. 1980). However, if $n$ is not large enough for the normality assumption to hold via the Central Limit Theorem, then (a) the Mahalanobis distance will not be $\chi_p^2$, and (b) the independence in Lemma 1 will  not hold. To address (a), the empirical distribution of each $M_f$ can be used instead of the $\chi^2_p$ distribution to select each corresponding threshold $a_f$. As for (b), the elements of $\{\bar{{\bm{x}}}_{f^+} - \bar{{\bm{x}}}_{f^-} : f \in F \}$ (and, as a consequence, the elements of $\{M_f: f \in F\}$) are always uncorrelated under our proposed rerandomization algorithm. This implies that, under mild regularity conditions, rerandomization will still increase the precision of factorial effect estimators; however, theoretical results found in Theorems 3 and 4 will not hold exactly.

\section{Illustrating Rerandomization in a $2^5$ Education Example} \label{s:schools}

Dasgupta, et al. (2015) discuss an educational experiment planned by the New York Department of Education (NYDE) with five ``incentive programs'' to be introduced to high schools ``which desperately need performance improvement.'' The programs include a quality review, a periodic assessment, inquiry teams, a school-wide performance bonus program, and an online-resource program (Dasgupta, et al. 2015).

The NYDE measures schools' performance with a score in each school's \textit{Progress Report}, and we consider nine covariates that will likely be correlated with this score: Total number of students, five different race variables (proportion of white, black, Asian, Native American, and Latino students), proportion of female students, enrollment rate, and poverty rate. This situation can be considered an extreme case of a ``tiers of covariates'' framework, where a subset of nine covariates are considered ``important'' and the rest are considered ``not important.'' The goal is to assign 43 schools to each of the 32 different treatment combinations such that the factorial effects of the experiment will be well-estimated.

Interest usually focuses on main effects and possibly two-way interactions, and higher-order interactions are often considered negligible (Wu and Hamada 2009). Thus, we implement a rerandomization algorithm that considers main effects ``most important,'' two-way interactions ``less important,'' and higher-order interactions ``not important.'' We created fifteen squared Mahalanobis distances: one for each of the five main effects and ten two-way interactions. The rerandomization algorithm involves randomizing until $\max (M_1, \dots, M_5) \leq a_{\text{main}}$ and $\max (M_6, \dots, M_{15}) \leq a_{\text{interaction}}$, where $a_{\text{main}}$ is the $100(0.01^{1/5})$ percentile of the $\chi^2_9$ distribution so $P(M_1, \dots, M_5 \leq a_{\text{main}}) = 1\%$, because, according to Lemma 1, the squared Mahalanobis distances are independent (and approximately $\chi^2_9$). Similarly, $a_{\text{interaction}}$ is the $100(0.1^{1/10})\%$ percentile of the $\chi^2_9$ distribution, making the criterion corresponding to the interaction effects less stringent than that of the main effects.

\begin{figure}[h!]
	\centering
	\includegraphics[scale = 0.75]{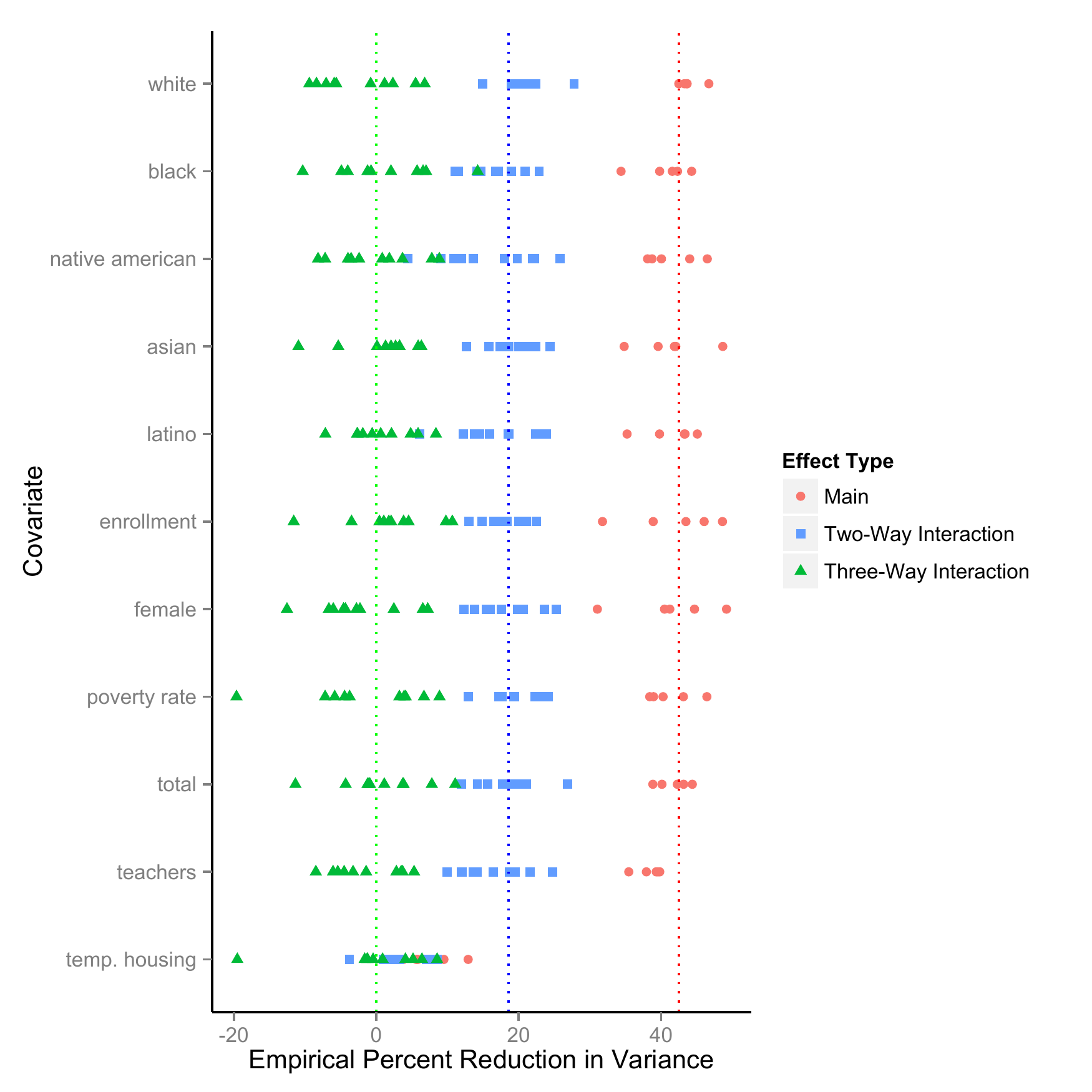}
	\caption{Percent reduction in variance in the covariate mean difference after rerandomization for various covariates and factorial effects. The expected percent reduction in variance given Theorem 3 for each type of factorial effect is marked by a vertical line. Displayed are the nine covariates considered during rerandomization as well as ``number of teachers'' and ``number of students in temporary housing,'' which were not considered.}
	\label{fig:lovePlot}
\end{figure}

We performed pure randomization and rerandomization 1,000 times. For each (re)randomization, the covariate mean difference $(\bar{x}_{p, f^+} - \bar{x}_{p, f^-})$ was calculated for each covariate $p$ and factor/interaction $f$. Figure \ref{fig:lovePlot} displays the empirical percent reduction in variance, which shows how much rerandomization reduced the variance of the covariate mean difference for various covariates and factors/interactions compared to pure randomization. Main effects are marked with circles, two-way interaction effects with squares, and three-way interaction effects with triangles. The percent reduction in variance expected given Theorem 3 is marked by a vertical line for each type of factorial effect.

The nine covariates we considered during rerandomization are displayed at the top of the vertical axis of Figure \ref{fig:lovePlot}. Rerandomization reduced the variance of the covariate mean difference across factors and two-way interactions compared to pure randomization for these covariates, and the reduction varies around what we would expect given Theorem 3. There is more reduction for individual factors than for interactions, as is expected, because the threshold $a_{\text{main}}$ was more stringent than $a_{\text{interaction}}$. The percent reduction in variance across three-way interactions is occassionally negative - implying randomization yielded better covariate balance in this case - but this reduction averages close to zero, as expected. Therefore, rerandomization on average increased the covariate balance across main effects and two-way interactions without sacrificing the covariate balance across higher-order interactions. Although one may be concerned about some fairly negative values for three-way interactions - there are two percent reduction in variances below -19\% - this behavior is similar to what would happen if we instead compared 1,000 randomizations to 1,000 different randomizations. On average, rerandomization was equivalent to randomization in terms of three-way interactions, which is expected, because three-way interactions were not considered during rerandomization.

Figure \ref{fig:lovePlot} also displays the percent reduction in variance for two covariates not considered during rerandomization: ``number of teachers'' and ``number of students in temporary housing.'' Rerandomization yielded more balance for ``number of teachers'' compared to pure randomization, because ``number of teachers'' is highly correlated ($R^2 = 0.95$) with ``number of students,'' which was considered during rerandomization. Likewise, ``number of students in temporary housing'' was only mildly correlated with the covariates considered during rerandomization, and thus it did not benefit greatly from rerandomization. If the NYDE decided that these two covariates were important to balance, but less so than the nine covariates already considered, we could rerandomize efficiently by balancing only the functions of ``number of teachers'' and ``number of students in temporary housing'' that are orthogonal to the nine ``most important'' covariates, because the parts that are correlated will already be balanced.

If outcome variables of the NYDE experiment are correlated with these covariates, then rerandomization will yield more precise estimates of the main factorial effects and two-way interactions. Furthermore, the precision of higher-order factorial effects will not be worse compared to pure randomization.

\section{Conclusion}

Rerandomization is known to increase the precision of the treatment effect estimator for treatment-versus-control experiments (Morgan and Rubin 2012). Here, rerandomization has been explored for balanced $2^K$ factorial designs. Theoretical results under common assumptions show that rerandomization yields unbiased estimators and increases the precision of factorial effect estimators of interest without sacrificing the precision of other estimators. Empirical results illustrate these theoretical results via a real-data application. The rerandomization algorithm can also be adjusted so tiers of importance for covariates and factorial effects can be incorporated. Extensions for more complex designs - such as unbalanced designs and fractional factorial designs - will be future work. 

\section{Appendix}

\noindent
\textit{Proof of Lemma 1}

\noindent
Assume a completely randomized balanced $2^K$ factorial design is rerandomized using the algorithm proposed at the end of Section \ref{ss:obs}, and the covariate means are multivariate normal. Under both randomization and rerandomization, the columns of $\widetilde{\bm W}$ defined in (\ref{eq:Wf}) are orthogonal. Because the factorial design is balanced and the criterion function $\phi$ defined in (\ref{eq:phi}) is symmetric in $\widetilde{\bm W}$, $\mathbb{E}[\widetilde{\bm W}_{.f} | \phi({\mathbf{X},\widetilde{\bm{W}}}) = 1] = \mathbf{0}$ for all $f \in F$. Therefore, for any $f_1, f_2 \in F$, $\text{cov}({\widetilde{\bm{W}}}_{f_1}, {\widetilde{\bm{W}}}_{f_2} | \phi({\mathbf{X},\widetilde{\bm{W}}}) = 1) = \mathbf{0}$.

Therefore, $\text{Cov}(\widetilde{\bm{W}} | \phi({\mathbf{X},\widetilde{\bm{W}}}) = 1)$ is a block-diagonal matrix. Because $\bar{{\bm{x}}}_{f^+} - \bar{{\bm{x}}}_{f^-} = \frac{\mathbf{X}^T \widetilde{\bm W}_{.f}}{n/2}$, the covariance matrix of the elements of $\{ \bar{{\bm{x}}}_{f^+} - \bar{{\bm{x}}}_{f^-} : f \in F \}$ is block-diagonal under rerandomization. By assumption the covariate means are multivariate normal, and thus this block-diagonal covariance matrix implies the elements of $\{\bar{{\bm{x}}}_{f^+} - \bar{{\bm{x}}}_{f^-}: f \in F\}$ are mutually independent under rerandomization. Additionally, the elements of $\{M_f: f \in F \}$ are mutually independent, because every $M_f$ is a function of $\bar{{\bm{x}}}_{f^+} - \bar{{\bm{x}}}_{f^-}$. Similarly, the elements of $\{\phi_f({\mathbf{X}, \widetilde{\bm{W}}}): f \in F\}$ are mutually independent, where $\phi_f({\mathbf{X}, \widetilde{\bm{W}}})$ is defined in (\ref{eq:phif}). $\hspace{0.1 in} \square$ \\
\noindent
\textit{Proof of Theorem 3}

\noindent
Assume a completely randomized balanced $2^K$ factorial design is rerandomized using the algorithm proposed at the end of Section \ref{ss:obs}, and the covariate means are multivariate normal. The elements of $\{\phi_f({\mathbf{X}, \widetilde{\bm{W}}}): f \in F\}$ are mutually independent given Lemma 1. Therefore,
\begin{align*}
	\mathbb{E}[\overline{{\bm{x}}}_{f^+} - \overline{{\bm{x}}}_{f^-} | \phi({\mathbf{X},\widetilde{\bm{W}}}) = 1] &= \mathbb{E}[\overline{{\bm{x}}}_{f^+} - \overline{{\bm{x}}}_{f^-} | \phi_f({\mathbf{X}, \widetilde{\bm{W}}}) = 1] \\
	&= \mathbb{E}[\overline{{\bm{x}}}_{f^+} - \overline{{\bm{x}}}_{f^-} | M_f \leq a]
\end{align*}
where $\phi({\mathbf{X},\widetilde{\bm{W}}})$ is defined in (\ref{eq:phi}). Similarly, for $f_1 = f_2$,
\begin{align*}
	\text{cov}[\overline{{\bm{x}}}_{f_1^+} - \overline{{\bm{x}}}_{f_1^-}, \overline{{\bm{x}}}_{f_2^+} - \overline{{\bm{x}}}_{f_2^-}| \phi({\mathbf{X},\widetilde{\bm{W}}}) = 1] &= \text{cov}[\overline{{\bm{x}}}_{f_1^+} - \overline{{\bm{x}}}_{f_1^-}, \overline{{\bm{x}}}_{f_2^+} - \overline{{\bm{x}}}_{f_2^-} | \phi_f({\mathbf{X}, \widetilde{\bm{W}}}) = 1] \\
	&= \text{cov}[\overline{{\bm{x}}}_{f_1^+} - \overline{{\bm{x}}}_{f_1^-}, \overline{{\bm{x}}}_{f_2^+} - \overline{{\bm{x}}}_{f_2^-} | M_f \leq a]
\end{align*}
while for $f_1 \neq f_2$,
\begin{align*}
	\text{cov}[\overline{{\bm{x}}}_{f_1^+} - \overline{{\bm{x}}}_{f_1^-}, \overline{{\bm{x}}}_{f_2^+} - \overline{{\bm{x}}}_{f_2^-}| \phi({\mathbf{X},\widetilde{\bm{W}}}) = 1] = \mathbf{0}
\end{align*}
because the elements of $\{ \bar{{\bm{x}}}_{f^+} - \bar{{\bm{x}}}_{f^-} : f \in F \}$ are mutually independent. The remainder of the proof is identical to the treatment-versus-control case, where the units assigned to the high level of $f$ are the ``treatment'' and the units assigned to the low level are the ``control.'' Thus, analogous to Morgan and Rubin (2012), for $f_1 = f_2$,
\begin{align*}
	\text{cov}[\overline{{\bm{x}}}_{f_1^+} - \overline{{\bm{x}}}_{f_1^-}, \overline{{\bm{x}}}_{f_2^+} - \overline{{\bm{x}}}_{f_2^-} | M_f \leq a] = v_a \text{cov}[\overline{{\bm{x}}}_{f_1^+} - \overline{{\bm{x}}}_{f_1^-}, \overline{{\bm{x}}}_{f_2^+} - \overline{{\bm{x}}}_{f_2^-}]
\end{align*}
where $v_a$ is defined as in (\ref{eq:va}). \hspace{0.1 in} $\square$ \\
\noindent
\textit{Proof of Theorem 4}

\noindent
Assume (a) a completely randomized balanced $2^K$ factorial design is rerandomized using the algorithm proposed at the end of Section \ref{ss:obs}, (b) the covariate means are multivariate normal, (c) factorial effects are constant across units, and (d) there is no interaction between factorial effects and covariate effects. Because the factorial effects are constant across units, each factorial effect estimator $\hat{\theta}_f$ can be written as (\ref{eq:linearModel}). By Lemma 1, for $f_1 \neq f_2$, $\text{cov}(\overline{{\bm{x}}}_{f_1^+} - \overline{{\bm{x}}}_{f_1^-}, \overline{{\bm{x}}}_{f_2^+} - \overline{{\bm{x}}}_{f_2^-} | \phi({\mathbf{X},\widetilde{\bm{W}}}) = 1) = \mathbf{0}$. Furthermore, the difference of the covariate means is orthogonal to the difference of the residual means, and therefore the covariance between them is zero. Therefore,
\begin{align*}
	\text{cov}(\hat{\theta}_{f_1}, \hat{\theta}_{f_2} | \phi({\mathbf{X},\widetilde{\bm{W}}}) = 1) = \text{cov}[(\bar{\epsilon}_{f_1^+} - \bar{\epsilon}_{f_1^-}) , (\bar{\epsilon}_{f_2^+} - \bar{\epsilon}_{f_2^-}) | \phi({\mathbf{X},\widetilde{\bm{W}}}) = 1] = 0 
\end{align*}
The final equality holds because, by the balance of the design, under both randomization and rerandomization,
\begin{align*}
	\text{cov}(\bar{\epsilon}_{f_1^+}, \bar{\epsilon}_{f_2^+}) = \text{cov}(\bar{\epsilon}_{f_1^+}, \bar{\epsilon}_{f_2^-}) = \text{cov}(\bar{\epsilon}_{f_1^-}, \bar{\epsilon}_{f_2^+}) = \text{cov}(\bar{\epsilon}_{f_1^-}, \bar{\epsilon}_{f_2^-})
\end{align*}
and thus the covariance between any two factorial effect estimators under rerandomization is zero. Furthermore, for all $f \in F$,
\begin{align*}
	\text{var}(\hat{\theta}_f | \phi({\mathbf{X},\widetilde{\bm{W}}}) = 1) &= \boldsymbol{\beta}^T \text{cov}(\bar{{\bm{x}}}_{f^+} - \bar{{\bm{x}}}_{f^-} | \phi({\mathbf{X},\widetilde{\bm{W}}}) = 1)\boldsymbol{\beta} + \text{var}(\bar{{\bm{\epsilon}}}_{f^+} - \bar{{\bm{\epsilon}}}_{f^-} | \phi({\mathbf{X},\widetilde{\bm{W}}}) = 1) \\
	&= v_a \boldsymbol{\beta}^T \text{cov}(\bar{{\bm{x}}}_{f^+} - \bar{{\bm{x}}}_{f^-})\boldsymbol{\beta} + \text{var}(\bar{{\bm{\epsilon}}}_{f^+} - \bar{{\bm{\epsilon}}}_{f^-} | \phi({\mathbf{X},\widetilde{\bm{W}}}) = 1) \\
	&= v_a \boldsymbol{\beta}^T \text{cov}(\bar{{\bm{x}}}_{f^+} - \bar{{\bm{x}}}_{f^-})\boldsymbol{\beta} + \text{var}(\bar{{\bm{\epsilon}}}_{f^+} - \bar{{\bm{\epsilon}}}_{f^-})
\end{align*}
The second equality is a result of Theorem 3. By assumption $n$ is large enough that $\bar{{\bm{x}}}_{f^+} - \bar{{\bm{x}}}_{f^-}$ and $\bar{{\bm{\epsilon}}}_{f^+} - \bar{{\bm{\epsilon}}}_{f^-}$ are normally distributed, and thus orthogonality implies independence. Thus, rerandomization does not affect the variance of $\bar{{\bm{\epsilon}}}_{f^+} - \bar{{\bm{\epsilon}}}_{f^-}$, and the final equality holds. The remainder of the proof is analogous to Morgan and Rubin (2012), because it is identical to the treatment-versus-control case, as in the proof of Theorem 3. \hspace{0.1 in} $\square$

\end{document}